%Paper: hep-th/9409064
%From: "Igor Klebanov" <klebanov@puhep1.Princeton.EDU>
%Date: Mon, 12 Sep 94 11:48:11 -0400

\input phyzzx

\catcode`\@=11
\def\half{{1\over 2}}
\def\CO{{\cal O}}
\def\eqaligntwo#1{\null\,\vcenter{\openup\jot\m@th
\ialign{\strut\hfil
$\displaystyle{##}$&$\displaystyle{{}##}$&$\displaystyle{{}##}$\hfil
\crcr#1\crcr}}\,}
\catcode`\@=12

\def\qg{quantum gravity}

\def\gb{\bar\gamma}
\def\L{Liouville}
\def\Tr{{\rm Tr}\,}

\def\mm{matrix model}
\def\NP{{\it Nucl. Phys.\ }}

\def\PL{{\it Phys. Lett.\ }}
\def\PR{{\it Phys. Rev.\ }}
\def\PRL{{\it Phys. Rev. Lett.\ }}
\def\CMP{{\it Comm. Math. Phys.\ }}

\def\IJMP{{\it Int. Jour. Mod. Phys.\ }}
\def\Mod{{\it Mod. Phys. Lett.\ }}

\REF\MAT{V. Kazakov, \PL {\bf 150B} (1985) 282;
J.~Ambj\o rn, B.~Durhuus, and J.~Fr\"ohlich \NP
{\bf B257 }  (1985) 433;
F. David, \NP {\bf B257} (1985) 45; V. Kazakov, I. Kostov
and A. Migdal, \PL {\bf 157B} (1985) 295.}
\REF\GM{
D.~J.~Gross and A.~A.~Migdal, \PRL {\bf 64} (1990) 717;
M.~Douglas and S.~Shenker, \NP {\bf B335} (1990) 635;
E.~Brezin and V.~Kazakov, \PL {\bf 236B} (1990) 144.}
\REF\sasha{A. M. Polyakov, \PL {\bf 103B} (1981) 207, 211.}
\REF\kpz{V. Knizhnik, A. Polyakov and A. Zamolodchikov,
\Mod {\bf A3} (1988) 819.}
\REF\thorn{T. L. Curtright and C. B. Thorn, \PRL {\bf 48} (1982) 1309. }
\REF\ddk{F. David, \Mod {\bf A3} (1988) 1651;
J. Distler and H. Kawai, \NP {\bf B321} (1989) 509.}
\REF\ns{N. Seiberg, {\it Prog. Theor. Phys. Suppl.} {\bf 102} (1990) 319;
N. Seiberg and S. Shenker, \PR {\bf D45} (1992) 4581.}
\REF\Mig{
M. Agishtein and A. A. Migdal,
Nucl.~Phys.~{\bf B350}, 690 (1991). }
\REF\Jain{S. Jain and S. Mathur, \PL {\bf B286} (1992) 239. }
\REF\Kawai{H.~Kawai, N.~Kawamoto, T.~Mogami, and Y.~Watabiki,
Phys.~Lett.~{\bf B306} (1993) 19;
S. S. Gubser and I. R. Klebanov, \NP {\bf B416} (1994) 827. }
\REF\Das{
S.~R.~Das, A.~Dhar, A.~M.~Sengupta, and S.~R.~Wadia,
\Mod {\bf A5} (1990) 1041. }
\REF\abc{ L. Alvarez-Gaum\' e, J.~L. Barbon and C. Crnkovic,
\NP {\bf B394} (1993) 383. }
\REF\korch{ G.~Korchemsky,
\Mod {\bf A7} (1992) 3081; \PL {\bf 296B} (1992)
323. }
\REF\sug{F. Sugino and O. Tsuchiya, UT-Komaba preprint 94-4,
hep-th 9403089.}
\REF\SI{S. Gubser and I. R. Klebanov, Princeton preprint PUPT-1479,
hep-th 9407014, to appear in Physics Letters B.}
\REF\I{I. R. Klebanov, Princeton preprint PUPT-1486,
hep-th 9407167.}
\REF\dur{B. Durhuus, preprint KUMI-94-2, hep-th 9402052;
J. Ambjorn, preprint NBI-HE-93-31, hep-th 9408129;
J. Ambjorn, B. Durhuus and T. Jonsson, \Mod {\bf A9} (1994) 1221.}
\REF\KSB{I. R. Klebanov, L. Susskind and T. Banks,
\NP {\bf B317} (1989) 665.}
\REF\BPIZ{E. Brezin, C. Itzykson, G. Parisi and J. Zuber,
\CMP {\bf 59} (1978) 35.}
\REF\Boul{D. Boulatov and V. Kazakov, \PL {\bf B186} (1987) 379. }
\REF\GK{D. J. Gross and I. R. Klebanov, \NP {\bf B344} (1990) 475.}
\REF\GKN{D. J. Gross, I. R. Klebanov and M. J. Newman
\journal Nucl. Phys. &B350 (1991) 621.}
\REF\BK{M. Bershadsky and I. R. Klebanov, \PRL {\bf 65} (1990) 3088.}
\REF\st{N. Sakai and Y. Tanii, \IJMP {\bf A6} (1991) 2743.}
\REF\BKN{M. Bershadsky and I. R. Klebanov, \NP {\bf B360} (1991) 559.}
\REF\kkd{V. Kazakov, I. Kostov and J.-M. Daul, \NP {\bf B409} (1993) 311.}
\REF\lp{A. Larkin and S. Pikin, {\it ZhETF} {\bf 56} (1969) 1664. }

\nopagenumbers
{\baselineskip=16pt
\line{\hfil PUPT-1498}
\line{\hfil September 1994}
\line{\hfil {\tt hep-th/9409064}}
 }

\overfullrule = 0 pt
\normalbaselineskip  = 18pt plus 0.2pt minus 0.1pt
\hsize = 6.2 in
\vsize = 8.9 in
\hoffset =  0.25 in
\title
{NON-PERTURBATIVE SOLUTION OF MATRIX MODELS MODIFIED BY
TRACE-SQUARED TERMS
}
\author {Igor R. Klebanov and Akikazu Hashimoto
}
\address{\JHL}

\abstract
We present a non-perturbative solution of
large $N$ matrix models modified by terms of the form
$ g(\Tr\Phi^4)^2$, which add microscopic wormholes to the
random surface geometry. For $g<g_t$ the sum over surfaces is in the
same universality class as the $g=0$ theory, and
the string susceptibility exponent is reproduced by
the conventional Liouville interaction $\sim e^{\alpha_+ \phi}$.
For $g=g_t$ we find a different universality class, and the string
susceptibility exponent agrees for any genus with Liouville theory
where the interaction term is dressed by the other branch,
$e^{\alpha_- \phi}$. This allows us to define a double-scaling limit
of the $g=g_t$ theory. We also consider matrix models modified by terms
of the form $g O^2$, where $O$ is a scaling operator.
A fine-tuning of $g$ produces a change in this operator's gravitational
dimension which is, again, in accord with the change in the branch of
the Liouville dressing.

\endpage

\pagenumbers
\chapter{Introduction}

Large $N$ matrix models [\MAT] have proven to be a remarkable source of
information about two-dimensional \qg\ coupled to conformal matter with
$c\leq 1$. They are the only available method for calculating sums over
geometries to all orders in the genus expansion [\GM].
Some matrix model results have been reproduced directly in
\L\ gravity [\sasha-\ddk],
which gives us confidence that the discretized and
continuum approaches describe the same theory.

The \mm s which generate conventional discretized random surfaces
are formulated with only single-trace terms in the action, such as $\Tr
V(\Phi)$. Such models have been studied thoroughly, and their \L\
formulation is believed to be understood quite well. In gravitational
dressing of any operator one encounters a two-fold ambiguity associated
with the choice of branch of square root. For
agreement with the conventional \mm s this ambiguity is always resolved
by picking the branch which is smoothly connected with the semiclassical
limit $c\to -\infty$. Therefore, many results here are qualitatively
semiclassical [\ns].

In this paper we study a different class of \mm s whose action, in
addition to the single-trace terms, also contains trace-square terms
such as $ g(\Tr\Phi^4)^2$.
Terms of this kind can glue a pair of random surfaces together at a
plaquette.  This contact may be thought of as a tiny neck (a wormhole), so that
the network of touching surfaces may be assigned an overall genus and
overall area.
It is known that such microscopic wormholes are already abundant in the
conventional theory with $g=0$ [\Mig-\Kawai].
By changing $g$ we are essentially
changing the weight of some singular geometries in the path integral.
It is not surprising, therefore, that a small increase in $g$ does not change
the universal properties of the model. If, however, we fine tune
$g$ to a finite positive value $g_t$, then the universality class
of the large area phase transition changes. This phenomenon was first
observed [\Das] in a modified one-matrix model.
It was found that there exists a
critical value $g_t$ such that, for $g<g_t$, the large area behavior
of the sum over genus zero surfaces
gives the string susceptibility exponent $\gamma=-1/2$ characteristic
of pure gravity, \ie
$$ F_0 (A) \sim A^{-3+\gamma}
\sim A^{-7/2 } \ .
$$
For $g>g_t$ one finds branched polymer behavior with
$\gamma=1/2$, which is not very interesting because it corresponds to
degenerate world sheets.
Most interestingly, for $g=g_t$ there exists
a new type of critical behavior with string susceptibility exponent $1/3$.
This is the first example of a \mm\ where new critical behavior
occurs due to fine-tuned wormhole weights.

Since ref. [\Das] a number of other such modified \mm s have been
studied [\abc-\I]. In general, as the trace-squared coupling is increased to a
critical value $g_t$, the string susceptibility exponent jumps from
some negative value $\gamma$, found in a conventional \mm, to
a positive value
$$\gb={\gamma\over\gamma-1}\ . \eqn\newgamma$$
Essentially equivalent results have been obtained without using \mm s,
on the basis of direct combinatorial analysis [\dur]. For a long time the
positive values of string susceptibility exponent seemed very puzzling.
Recently, however, a simple continuum explanation of these critical
behaviors was proposed in ref. [\I].

For all the conventional \mm s describing $(p, q)$ minimal
models coupled to gravity, the correct scaling follows from the
\L\ interaction of the form
$$\eqalign{&\Delta \int d^2\sigma O_{\min} e^{\alpha_+\phi}\ ,\cr
&\alpha_+=
{1\over 2\sqrt 3}(\sqrt {1-c+24h_{\min}}-
\sqrt{25-c})=-{p+q-1\over \sqrt{2pq}}
\cr}
\eqn\usint$$
where $O_{\min}$ is the matter primary field of the lowest dimension,
$$h_{\min}={1-(p-q)^2\over 4pq}\ .\eqn\ldim$$
A simple calculation reveals that the
string susceptibility exponent is given by
$$\gamma= 2+ {Q\over\alpha_+}\ ,\eqn\geng$$
where
$ Q=\sqrt{25-c\over 3} $.
In ref. [\I] it was argued that the effect of fine-tuning the
touching interaction is to replace the \L\ potential by
$$\eqalign{&\Delta \int d^2\sigma O_{\min} e^{\alpha_-\phi}\ ,\cr
&\alpha_-=
-{1\over 2\sqrt 3}(\sqrt {1-c+24h_{\min}}+
\sqrt{25-c})=-{p+q+1\over \sqrt{2pq}} \ .
\cr}
\eqn\modpot$$
Now the string susceptibility exponent is found to be
$$\gb= 2+ {Q\over\alpha_-}={\gamma\over \gamma-1}\ .\eqn\newg$$
This establishes correspondence with the \mm\ results, eq.
\newgamma.
Thus, in the \L\ description of the modified \mm s we simply have to pick
the other branch of square root in gravitational dressing compared to
the conventional \mm s. This proposal has a number of interesting
implications that are worth studying.
For instance, using the scaling arguments of
ref. [\ddk] we find that the modified sum over surfaces of genus $G$
should obey the scaling law
$${\partial^2 \bar F_G\over \partial\Delta^2}
\sim {1\over \Delta^{2G+\gb (1-G)}}
\ .\eqn\scaling$$
In the \mm s this has only been checked for $G=0$ and 1.
If true for all $G$, eq. \scaling\ implies that it should be possible
to define a double scaling limit for the modified \mm s.

\section{Summary of Results}

In this paper we carry out a non-perturbative study of various
modified \mm s and confirm that eq. \scaling\ is indeed valid.
The simplest class of models we investigate are the modified
multicritical
one-matrix models [\abc,\korch],
$$ Z_k=\int {\cal D} \Phi \, e^{-N \,
   \left[ \Tr V_k(\Phi) + (c_2 -\lambda) \Tr \Phi^4-
    {g \over 2 N} \left( \Tr \Phi^4 \right)^2 \right]} \ .    \eqn\kth
$$
The critical potential of the $k$-th model with $g=0$ is
$$ V_k(\Phi)= \sum_{i=1}^k (-1)^{i+1} c_i\Phi^{2i}
\ ,$$
where $c_i$ have been determined in ref. [\GM]. We choose to study
the dependence  of the sum over surfaces on a deformation of the
potential by a term $\sim \Phi^4$.
For $g=0$, it is known that $\gamma=-1/k$.
The universal (leading singular) part of the sum
over connected surfaces, $F=\log Z$, is a function of the scaling variable
$t\sim (c_2-\lambda) N^{2/(2-\gamma)}$.
In fact, for all $g<g_t$ the sum over surfaces is in the same universality
class.
For the fine-tuned value $g=g_t$ the
string susceptibility exponent jumps to $\gb=1/(k+1)$.
At this point there exists a critical value $\lambda_c$ such that
the universal part of the sum over connected surfaces, $\bar F$,
is a function of the scaling variable $\bar t\sim (\lambda_c-\lambda)
N^{2/(2-\gb)}$. Our calculations reveal a remarkably simple
non-perturbative relation
$$\bar F (\bar t) =\log \int_{-\infty}^\infty dt
e^{t \bar t + F(t)}
\eqn\double$$
which connects the double scaling limit of modified \mm s with that
of the conventional ones. We will show that this relation is quite
general; it applies to all modified one-matrix and two-matrix models
with trace-squared terms of the simplest kind, which describe $c<1$
models coupled to gravity. \foot{For
$c=1$ the relation is somewhat different due to the logarithmic
scaling violations.}

There is a way, however, to alter
the relation \double\ by changing the type of trace-squared terms
added to the action. We may replace the term which is the square
of the lowest dimension operator by the square of some other scaling
operator $O$. Adding $g O^2$ to the action and
 fine-tuning $g$, we obtain a model where
the gravitational dimension of the operator $O$ changes from $d$ to
$$\bar d=\gamma-d\eqn\newgd$$
(the string susceptibility exponent remains
unchanged). Remarkably, this change of
dimension is again reproduced in \L\ theory by a mere
change of the branch of gravitational dressing. Namely, if the operator
is dressed by $e^{\beta_\pm\phi}$ then
$$ d= 1- {\beta_+\over \alpha}, \qquad\qquad
\bar d= 1- {\beta_-\over \alpha}
$$
where $e^{\alpha\phi}$ is the dressing of the \L\ potential.
The relation \newgd\ follows from $\beta_++\beta_-= -Q$.
This gives us some further evidence in favor of the interpretation in
ref. [\I].

In the \mm s we, in fact, find the non-perturbative dependence on the
coupling constant corresponding to $O$.
If we perturb the action of the model
by a term $\tau_O O$ then, for $g=0$, the universal free energy $F$ is
a function of $t$ and $t_O= \tau_O N^{2(1-d)/(2-\gamma)}$.
For $g=g_t$ we obtain a new universal free energy
$$\bar F (t,\bar t_O) =\log \int_{-\infty}^\infty dt_O
e^{t_O \bar t_O + F(t, t_O)}
\eqn\newdouble$$
where $\bar t_O\sim \tau_O N^{2(1-\bar d)/(2-\gamma)}$.
Thus, the calculation reduces to an integral over a different coupling
constant from that in eq. \double. Now it is clear that, by
simultaneously adding $n$
different types of trace-squared terms and fine tuning their coefficients
we can change the gravitational dimensions of $n$ operators in the
theory. The resulting partition function is a transform of the original
partition function with respect to corresponding coupling constants
$t_i$. The most general relation is
$$\bar F (\{\bar t\}, \{ T \}) =\log \prod_{i=1}^n\int_{-\infty}^\infty dt_i
e^{\sum_{j=1}^n t_j \bar t_j + F(\{ t\}, \{ T \})}
\eqn\gendouble$$
where $\{ T \}$ is some set of other coupling constant that remain
unintegrated.

The organization of the rest of the paper is as follows.
In section 2 we derive the integration over the coupling constants in eq.
\double\ via a trick familiar from wormhole physics.
In section 3 we demonstrate the change of operator gravitational
dimensions and derive eqs. \newdouble\ and \gendouble.
In section 4 we study the torus free energy which is directly related
to the operator content of the theory. We attempt to interpret the
modified \mm\ results in terms of \L\ theory.
We conclude in section 5.

\chapter{Wormholes and Integration over Coupling Constants}

In this section we derive non-perturbative results for the sum over
surfaces in modified \mm s.

\section{The one-matrix models}

Our method is completely general but
we will first illustrate it with the simplest example, the modified
$k=2$ one-matrix model which describes pure gravity.
We need to study the following integral over an $N\times N$
hermitian matrix $\Phi$,
$$ Z=\int {\cal D} \Phi \, e^{-N \,
   \left[ \Tr \left( \half\Phi^2 -
     \lambda \Phi^4 \right) -
    {g \over 2 N} \left( \Tr \Phi^4 \right)^2 \right]} \ .    \eqn\om
$$
A very helpful trick is to rewrite this as
$$ Z= {N\over \sqrt {2\pi g}}\int_{-\infty}^\infty dy e^{-{N^2 y^2\over 2g}}
\int {\cal D} \Phi \, e^{-N \,  \Tr \left(\half \Phi^2 -
(\lambda+y) \Phi^4 \right) }\ .
\eqn\trick$$
This trick is generally useful when the action is perturbed by
a square of an operator and has been applied extensively in wormhole
physics [\KSB]. In fact, the wormhole combinatorics leading to
eq. \trick\ is the same as in four dimensions.
Matrix models implement this combinatorics automatically, so that
eq. \trick\ can be derived in one line.
As in four dimensions we find integration over a coupling
constant, which in this case is the quartic coupling.
Here we are on a much firmer ground, however, because
we know a great deal about the
euclidean path integral, which is given by the logarithm of the
matrix integral. This will allow us to obtain perfectly
explicit and interesting results.

The advantage of representation \trick\ is that we know how to perform
the integral over $\Phi$.
The remaining integral over $y$ is one-dimensional and can be thought
of as an effective theory of baby universes, which lends itself to
perturbative expansion around the saddle point.
Relying on the well-known solution of the one-matrix model [\BPIZ, \GM],
we have
$$\eqalign{& \log \int {\cal D} \Phi \, e^{-N \, \Tr \left(\half \Phi^2 -
(\lambda+y) \Phi^4 \right)} =N^2(-a_1 x + \half a_2 x^2) + F(x, N^2) \ ,\cr
& F(x, N^2)= N^2 (-{2\over 5}a_3 x^{5/2}+\ldots )+ N^0 (-{1\over 24}\log
x+\ldots )+ N^{-2} (a_4 x^{-5/2}+\ldots) +\CO(N^{-4})\ , \cr
& x= c_2 - (\lambda+y) \ .}
\eqn\auxom
$$
We have chosen to separate the free energy into its singular part, $F$,
and the leading non-singular parts of order $N^2$ which play
an important role in our discussion.
{}From the leading order solution of ref. [\BPIZ] we know the
coefficients
$$ a_1=4\ ,\qquad a_2=576\ ,\qquad a_3=6144\sqrt 3\ ,\qquad c_2={1\over
48\ .}
$$
In the double scaling limit, $t=x N^{4/5} a_3^{2/5}$ is held fixed
so that the subleading parts of $F$ at each order in $N$ become
negligible.
Thus, in this limit
$$ F(x, N^2)= F(t)=-{2\over 5} t^{5/2} -{1\over 24} \log t+
{7\over 2160} t^{-5/2}+ \CO(t^{-10})
\eqn\dsf$$
where we neglected $x$-independent additive terms.
The expansion of $F(t)$ follows from the fact that
$\chi(t)= {d^2 F\over dt^2}$ satisfies the Painlev\' e equation [\GM]
$${d^2 \chi\over dt^2}=3(t-\chi^2)
\ .$$

{}From eqs. \trick\ and \auxom, we have
$$\eqalign{&Z={N\over \sqrt {2\pi g}}
 \int_{-\infty}^\infty dx e^{f(x)}\ ,\cr
& f(x)={N^2\over 2g}\left [(c_2-\lambda)^2+ 2x(c_2-a_1 g-\lambda)-
x^2 (1-a_2 g)\right ]+ F(x, N^2)
 \ . }
\eqn\nicerep$$
This may be thought of as an effective theory
of baby universes. The interaction vertex of
$n$ baby universes arises from surfaces with $n$
punctures (each puncture is
generated by insertion of operator
$\Tr\Phi^4$). The explicit quadratic term in $x$ gives
a kind of mass term (inverse propagator)
for the baby universe. Thus, the mass-squared is given by
$m^2={1\over g}- a_2$, with the first contribution coming from the wormhole
term in the matrix model, and the second from
the degenerate sphere consisting of two plaquettes.
We will analyze the three cases where the mass-squared is
positive, negative and zero separately.
The essential observation is that,
since $f(x)$ is of order $N^2$, we may develop a large $N$ expansion by
integrating around the saddle point $x_s$ given by
$f'(x_s)=0$.

Let us now show that, in
the massive case $g<1/a_2$ the sum over surfaces is in the
same universality class as the $g=0$ theory.
Defining
$$\Delta=\lambda_c-\lambda, \qquad\qquad \lambda_c=c_2-a_1 g  $$
we find that the location of the saddle point is
$$x_s ={\Delta\over 1-a_2 g }+\CO(\Delta^{3/2})
$$
Here the best way to analyze eq. \nicerep\ is by shifting the integration
variable, $z= x- {\Delta\over 1-a_2 g }$. Discarding some non-singular terms
in $\Delta$, we have
$$\log Z (\Delta, N^2)=\log
 \int_{-\infty}^\infty N dz
\exp \left [-{N^2(1-a_2 g)\over 2g}
z^2 + F\left(z+ {\Delta\over 1-a_2 g }, N^2\right) \right ]
\ .$$
After rescaling the variables,
$t={\Delta\over 1-a_2 g } N^{4/5} a_3^{2/5}\ ,\
\tilde z= z N^{4/5} a_3^{2/5} \ ,$
we arrive at
$$\log Z (\Delta, N^2)=\log
 \int_{-\infty}^\infty N^{1/5} d\tilde z
\exp \left [-{N^{2/5}a_3^{-4/5} (1-a_2 g)\over 2g}
\tilde z^2 + F(t+ \tilde z) \right ]\ .
$$
In the double-scaling limit, the gaussian term in the integrand becomes a
delta-function.
Therefore, the singular part of the
sum over surfaces satisfies
$$\log Z (\Delta, N^2)= F(t)\ ,
$$
where $F(t)$ is given in eq. \dsf.
This explicitly establishes the universality for $g<1/a_2$.
Thus, in their massive phase, the baby universes are irrelevant.
\foot{
In section 2.2 we show that this is true
is general. The massive baby universes are irrelevant due
to the general relation $d > \gamma/2$, where $d$ is the
gravitational dimension of the operator inserted by
the baby universe, and $\gamma$ is the string
susceptibility.  In the case just studied, $d=0$ and $\gamma = -1/2$.}

For the tachyonic case
$g>1/a_2$, the saddle point near $x=0$ becomes unstable, but
a stable saddle point appears at
$$ x_s= \tilde x+ \alpha\Delta^{1/2}+\CO(\Delta)\ , \qquad\qquad
\alpha>0\ ,
$$
where $\tilde x>0$ is determined by the equation $f''(\tilde x)=0$.
We also deduce that $f'(\tilde x)=N^2\Delta/g$.
Expanding $f(x_s)$ in powers of $\Delta$, we find that the
leading singularity in the planar limit is
$$\log Z (\Delta, N^2)\sim N^2 \Delta^{3/2}
$$
which is indicative of the branched polymer phase. This behavior
is quite generic because it does not depend on the precise form of
$F(x, N^2)$.

The massless case $g = 1/a_2$ is critical.
Here the position of the saddle point acquires a new scaling,
$$ x_s = \left ({\Delta a_2\over a_3}\right )^{2/3} +\CO(\Delta)\ . $$
Integrating around the saddle point in eq. \nicerep, we find
$$\log Z= \left [f-\half \log \left (-{f''\over N^2}\right )+
{ f^{(4)}\over 8 (f'')^2} - {5\over 24} {(f^{(3)})^2\over (f'')^3}
\right ]_{x=x_s}+\CO(1/N^4)
\ .\eqn\saddle$$
We find that all the terms in this expansion are important.
After some calculation, we arrive at
$$\eqalign{&\log Z= N^2 ({3\over 5}
\tilde\Delta^{5/3}+\ldots) + N^0(-{7\over 36}\log \tilde\Delta +\ldots)
+N^{-2} ({77\over 960} \tilde\Delta^{-5/3} +\ldots) +\CO(N^{-4})\ ,\cr
& \tilde\Delta= \Delta a_2 a_3^{-2/5}\ . }
$$
Thus, the singularity for any genus occurs as $\Delta\to 0$.
The structure of the leading singular terms suggests
that we may now define double scaling limit by keeping
the variable $\bar t\sim \Delta N^{6/5}$ fixed.
In fact, this follows directly from eq. \nicerep. For
$g=1/a_2$ the singular part
of the sum over surfaces, given by $\bar F=\log Z$, becomes
$$\bar F(\Delta, N^2)=\log
\int_{-\infty}^\infty dx N e^{N^2 a_2 \Delta x + F(x, N^2) }\ .
\eqn\mt$$
Introducing scaling variables
$$t=x N^{4/5} a_3^{2/5}\ ,\qquad\qquad \bar t = \Delta N^{6/5} a_2
a_3^{-2/5}\ , $$
we find that
$$\bar F(\Delta, N^2)=\bar F(\bar t)=
\log \int_{-\infty}^\infty dt
e^{t \bar t + F(t)}\ .
\eqn\main$$
This is our main result, which establishes a simple relation between the
double-scaling limits in the modified and conventional \mm s.
It is uncertain whether eq. \main\ has a truly non-perturbative meaning:
there are well-known problems in defining $F(t)$ non-perturbatively.
Even if they are overcome, it is not clear if the integral over $t$
will converge. What {\it is} certain is that eq. \main\ determines the
sum over modified surfaces of any genus, \ie\ it works to all orders
of perturbation theory.
Using saddle-point techniques, summarized in eqs.
\nicerep\ and \saddle, we may generate the large $\bar t$ expansion of
$\bar F$ directly from the integral representation \main\ and the
large $t$ expansion of $F$.

Eq. \main\ applies equally well to matrix models with non-symmetric
potentials, which are more basic because they do not double the
degrees of freedom. For such models we divide $F(t)$ from eq. \dsf\
by 2 to obtain the sum over surfaces. After a redefinition of $t$, we
find
$$F(t)=-{2\over 5} t^{5/2} -{1\over 48} \log t+
{7\over 8640} t^{-5/2}+ \CO(t^{-10}) \ .
$$
Substituting this into \main, and generating the saddle-point expansion,
we find the modified sum
over surfaces for non-symmetric models,
$$ \bar F(\bar t)= {3\over 5} \bar t^{5/3}-{13\over 72}\log \bar t+
{257\over 3840} \bar t^{-5/3} + \CO(\bar t^{-10/3})
$$
where we have explicitly calculated the contributions up to genus 2.

Extension of the methods presented above to the $k$-th multicritical
one-matrix model is quite straightforward.
First we rewrite eq. \kth\ as
$$ Z_k= {N\over \sqrt {2\pi g}}\int_{-\infty}^\infty dy e^{-{N^2 y^2\over 2g}}
\int {\cal D} \Phi \, e^{-N \,  \Tr \left( V_k(\Phi) +
(c_2-\lambda-y) \Phi^4 \right) }
\eqn\eq$$
Using the variable $x=c_2-\lambda-y$, we find
$$\eqalign{&Z_k={N\over \sqrt {2\pi g}}
 \int_{-\infty}^\infty dx e^{f_k(x)}\ ,\cr
& f_k(x)={N^2\over 2g}\left [(c_2-\lambda)^2+ 2x(c_2-a_1 g-\lambda)-
x^2 (1-a_2 g)\right ]+ F_k(x, N^2)\ ,\cr
& F_k(x, N^2)=-{k\over 2k+1} a_3 N^2 x^{(2k+1)/k}+\ldots }
\eqn\newrep$$
In the double scaling limit of the $g=0$ theory, $t\sim x N^{2k/(2k+1)}$
is held fixed, and the sum over surfaces is given by
$F_k(t)$.
Analysis of the saddle-point expansion shows that the theory is in the
same universality class for any $g<1/a_2$.

For $g=1/a_2$, we instead have
$$\eqalign{&\bar F_k(\Delta, N^2)=\log
\int_{-\infty}^\infty dx N e^{N^2 a_2 \Delta x + F_k(x, N^2) }\ ,\cr
&\Delta= c_2 -{a_1\over a_2}-\lambda\ . }
$$
Introducing scaling variables
$$t=x N^{2k/(2k+1)} a_3^{k/(2k+1)}\ ,\qquad\qquad
\bar t = \Delta N^{(2k+2)/(2k+1)} a_2
a_3^{-k/(2k+1)} $$
we arrive at the modified sum over surfaces in the double scaling limit,
$$\bar F_k(\bar t)=
\log \int_{-\infty}^\infty dt
e^{t \bar t + F_k(t)}
\ .\eqn\eq$$
{}From the fact that
the modified sum over surfaces is a function of $\bar t$, and that the
original expansion was in powers of $1/N^2$, it follows that the
genus $G$ contribution scales as $t^{(2k+1)(1-G)\over k+1}$.
Using the fact that, for non-symmetric matrix potentials,
$$ F_k (t) =-{k\over 2k+1} t^{(2k+1)/k} - {k-1\over 24 k}
 \log t + \sum_{j=1}^\infty \alpha_j(k)
t^{-j(2k+1)/k}
\eqn\multicrit$$
we generate the genus expansion of $\bar F_k$ with the saddle-point
methods,
$$\bar F_k(\bar t)={k+1\over 2k+1}\bar t^{2-{1\over k+1}}-
{1\over k+1} \left ({k-1\over 24} +{1\over 2}\right ) \log \bar t+
\CO(\bar t^{-2+{1\over k+1}})
\ .$$
This confirms the known result [\abc, \korch]
that on a sphere the string susceptibility
exponent is $\gb={1\over k+1}$. Thus, the order of the phase
transition for planar surfaces has changed from the third to the second.
We have also established
that the susceptibility exponent at genus $G$ is
$\gb + G(2-\gb)$, in agreement with eq.
\scaling. Since this genus dependence has such a natural
explanation in Liouville theory, it provides a solid argument in favor
of the Liouville interpretation of the modified \mm s [\I].

\section{The two-matrix models}

In this section we consider a more general class of \mm s and show that,
with the simplest kind of trace-squared terms, the non-perturbative
relation \double\ applies to them as well.

First we address the Ising model coupled to gravity, which is well known
to be described by a two-matrix model [\Boul]. Its simplest modified version
was introduced in ref. [\I],
$$ \eqalign{& Z_{\rm Ising}=\int {\cal D} \Phi_1 {\cal D} \Phi_2 \, e^{-N \,
   \left[ S(\Phi_1)+ S(\Phi_2)+ k\Tr \Phi_1 \Phi_2
    \right]} \ ,    \cr
& S(\Phi) =
   \Tr \left( \Phi^2 -
     \lambda \Phi^4 \right) -
    {g \over N} \left( \Tr \Phi^4 \right)^2\ . \cr }
\eqn\ising$$
We have added trace-squared terms of the form
$$ g [\left( \Tr \Phi_1^4 \right)^2 +\left( \Tr \Phi_2^4 \right)^2]
\eqn\modterms$$
which implies that the value of the Ising spin at the two ends of
a wormhole is required to be the same.
\foot{In the next section we will relax
this condition and allow a spin flip when a wormhole is traversed.
We will see
that more general theories can be constructed this way.}
Rewriting eq. \modterms\ as
$$ {g\over 2} \left( \Tr (\Phi_1^4+\Phi_2^4) \right)^2 +
{g\over 2}\left( \Tr (\Phi_1^4-\Phi_2^4) \right)^2
\eqn\tsq$$
and applying our trick to each of the two terms, we arrive at
$$ \eqalign{& Z_{\rm Ising}={N^2\over 2\pi g}
\int_{-\infty}^\infty dy dv e^{-{N^2 (y^2+v^2)\over 2g}}
\int {\cal D} \Phi_1 {\cal D} \Phi_2 \, e^{-N S}\ , \cr
& S =
   \Tr \left[ \Phi_1^2+\Phi_2^2+ k\Tr \Phi_1 \Phi_2-
     (\lambda+y) ( \Phi_1^4+ \Phi_2^4)- v ( \Phi_1^4- \Phi_2^4) \right]
\ .}\eqn\isingtrick$$
We perform the matrix integral first, and save the integrals over
$v$ and $y$ until the end.
The matrix integral describes the Ising model in magnetic field
$v$ [\Boul]. If we tune $k$ to its critical value and define $x=c_2-\lambda-y$,
where $c_2$ is the critical quartic coupling, then
$$\eqalign{& \log \int {\cal D} \Phi_1 {\cal D} \Phi_2 \, e^{-N S}
=N^2(-a_1 x + \half a_2 x^2) + F_0(x, N^2)+ F_1 (x, v, N^2) \ ,\cr
& F_0(x, N^2)= N^2 (-{3\over 7}a_3 x^{7/3}+\ldots )+ N^0 (-{1\over 12}\log
x+\ldots )+ \CO(N^{-2})\ , \cr
& F_1 (x, v, N^2)=\sum_{n=1}^\infty {v^{2n}\over (2n)!}
\VEV{\left( \Tr (\Phi_1^4-\Phi_2^4) \right)^{2n}}\ . }
\eqn\auxising
$$
The $v^{2n}$ vertex in the diagrammatic expansion for $v$ is given by the
connected correlation function of $2n$ operators $\Tr(\Phi_1^4 - \Phi_2^4)$.
This operator is the gravitationally dressed spin field, which is
known to have gravitational dimension $1/6$. Therefore,
$$
\VEV{\left( \Tr (\Phi_1^4-\Phi_2^4) \right)^2}= N^2 (b+ b' x^{2/3}+\ldots)
+ N^0 x^{-5/3}+ \ldots
$$
The inverse propagator (mass-squared) for $v$ is then given by
$ \left( {1\over g} - b \right) $.
By an explicit calculation, following the results of ref. [\Boul], we determine
$$ a_2 = {84024\over 625},\qquad\qquad b= {41256 \over 625}\ .
$$
It is important that $a_2 > b$. Let us imagine dialing $g$ up. For
$g<1/a_2$ both the $ x$ and $v$ baby universes are massive
and contribute only subleading terms to
the free energy. Here we find the same universality class as the unmodified
Ising model. For $g=1/a_2$, $x$ becomes massless and changes
the critical behavior. $v$ is still massive, however,
so that the integral over $v$
has the form
$$
\int_{-\infty}^\infty N dv \exp\left [-\half N^2 v^2 (a_2-b) +
F_1^{\rm universal} (x, v, N^2)\right ]
\ .$$
In terms of the rescaled variables
$\tilde v= v N^{2(1-d)/(2-\gamma)}$ and
$t\sim x N^{2/(2-\gamma)}$, this becomes
$$
\int_{-\infty}^\infty N^{{2d-\gamma\over 2-\gamma}} d\tilde v
\exp\left [-\half N^{{4d-2\gamma\over 2-\gamma}} \tilde v^2 (a_2-b) +
F_1 (t, \tilde v)\right ]
\ .$$
If $2d-\gamma>0$, then in the double scaling limit the gaussian
factor becomes a delta function, so that $\tilde v$ is frozen at zero.
\foot{This holds here because the gravitational dimension of the spin operator
is $d=1/6$ and $\gamma = -1/3$. In fact, this holds in general because in
all conventional \mm s $d\geq 0$ and $\gamma<0$. We conclude that
all massive baby universes are irrelevant. }
Since $F_1 (t, 0)=0$, the $\tilde v$ integral does not contribute to the
effective action for $x$.
Thus, all the dominant terms in $Z_{\rm Ising}$
can be calculated from
$$\eqalign{&Z_{\rm Ising}={N\over \sqrt {2\pi g}}
 \int_{-\infty}^\infty dx e^{f(x)}\ ,\cr
& f(x)={N^2\over 2g}\left [(c_2-\lambda)^2+ 2x(c_2-a_1 g-\lambda)-
x^2 (1-a_2 g)\right ]+ F_0(x, N^2) \ .}
\eqn\eq$$
{}From here on the calculation is analogous to those encountered
in the one-matrix model. Introducing scaling variables
$$\bar t \sim ( c_2 -{a_1\over a_2}-\lambda ) N^{8/7}\ ,\qquad\qquad
t\sim x N^{6/7}
$$
we arrive at the relation \double.

The calculation presented above for the Ising model can be carried over
almost verbatim to any modified two-matrix model of the form
$$ \eqalign{& Z=\int {\cal D} \Phi_1 {\cal D} \Phi_2 \, e^{-N \,
   \left[ S_p(\Phi_1)+ S_p(\Phi_2)+ k\Tr \Phi_1 \Phi_2
    \right]} \ ,    \cr
& S_p(\Phi) =
   \Tr \left( \Phi^2 -
     \lambda \Phi^4 +\ldots +\tau \Phi^{2p-2}\right) -
    {g \over  N} \left( \Tr \Phi^4 \right)^2\ . \cr }
\eqn\arbit $$
The parameters of the potential $S_p(\Phi)$ can be tuned
[\kkd] in such a way that this describes an arbitrary minimal
model coupled to gravity. For any such model, the modified
sum over surfaces $\bar F(\bar t) $ is related
to the conventional sum $F(t)$ by
eq. \double. The general validity of this relation implies the generality
of eq.  \newgamma. To show this, note that
$$\bar t \sim \Delta N^{2/(2-\bar\gamma)}\ ,\qquad\qquad
t\sim x N^{2/(2-\gamma)} \ .
$$
Since $t\bar t\sim \Delta x N^2$, we have
$$ {1\over 2-\bar\gamma}+ {1\over 2-\gamma} = 1 $$
from which eq.  \newgamma\ follows. Thus, if the asymptotic
expansion of $F(t)$ is in powers of $t^{2-\gamma}$, then the
asymptotic expansion of $\bar F(\bar t) $ is in powers of $\bar t^
{2-\bar\gamma}$.

\section{$c=1$}

One interesting theory that remains to be discussed is the $c=1$ model
coupled to gravity. We will consider compact target space of
radius $R$, which is described by matrix quantum mechanics at finite
temperature [\GK]. The path integral that generates the sum over touching
surfaces is [\sug,\SI]
$$
{\cal Z} = \int {\cal D} \Phi(t) \, e^{-N \int_{0}^{2\pi R} dt \,
   \left[ \Tr \left( {1 \over 2} \dot{\Phi}^2 + {1 \over 2} \Phi^2 -
     \lambda \Phi^3 \right) -
    {g \over 2N} \left( \Tr \Phi^3 \right)^2 \right]} \ ,
$$
with $\Phi(2\pi R)= \Phi(0)$.
Let us introduce
the normal mode operators
$$\eqalign {&P= \int_{0}^{2\pi R} dt \Tr \Phi^3(t) \ , \cr
&C_n ={1\over\sqrt 2} \int_{0}^{2\pi R} dt \cos{nt\over R} \Tr \Phi^3(t) \ ,
\cr
&S_n ={1\over\sqrt 2} \int_{0}^{2\pi R} dt \sin{nt\over R} \Tr \Phi^3(t)
\ , }
$$
and write the trace-squared term as a sum of squares
$$ \int_{0}^{2\pi R} dt \left( \Tr \Phi^3 \right)^2=
{1\over 2\pi R} \left ( P^2 +\sum_{n=1}^\infty (C_n^2+ S_n^2)\right )
$$
The operators $C_n$ and $S_n$ are known to have gravitational
dimension $d=n/2R$ [\GKN].
Using the by now familiar trick, we introduce a ``baby universe variable'' for
each squared operator in the action to derive
$$\eqalign {&
{\cal Z}\sim \int_{-\infty}^\infty dy_0 \prod_{n=1}^\infty dy_n dz_n
e^{-{\pi R N^2\over g}(y_0^2+y_n^2+z_n^2)}\cr
&\int {\cal D} \Phi(t) \, e^{-N \int_{0}^{2\pi R} dt \,
   \left[ \Tr \left( {1 \over 2} \dot{\Phi}^2 + {1 \over 2} \Phi^2 -
     (\lambda+y_0) \Phi^3 -\sum_{i=1}^\infty (y_i C_i+z_i S_i)\right)
\right]}   }
$$
Performing the matrix integral, we get
$$\eqalign{& \log \int {\cal D} \Phi(t) \, e^{-N S}
=2\pi R N^2(-a_1 x + \half a_2 x^2) +
F_0(x, N^2)+ F_1 (x, y_n, z_n, N^2) \ ,\cr
& F_0(x, N^2)=R N^2 (\half a_3 x^2/\log x+\ldots )-{1\over 24}
(R+{1\over R})\log
x+\ldots \cr
& F_1 (x, y_n, z_n, N^2)=\pi R N^2 \sum_{n=1}^\infty (y_n^2+z_n^2)
(b_n+ b_n' (x/|\log x|)^{n/R}+\ldots)+ \ldots }
\eqn\auxc
$$
We have exhibited the terms in $F_1$ that come from the two-point
functions of $C_n$ and $S_n$. These quadratic terms determine whether
the variables $y_n$ and $z_n$ become critical simultaneously with
the variable $x=c_2-\lambda-y_0$.

{}From a calculation of
the momentum dependence of the puncture two-point function [\GKN], we have
$$ b_n=\int_{-\infty}^\infty ds {s^2\over s^2+(n/R)^2} f^2(s)
$$
where $f(s)$ is proportional to the Fourier transform of the classical
trajectory at the top of the critical potential. Since $b_n$ is a
decreasing function of $n$, and $b_0=a_2$, we conclude that
$b_n< a_2$ for all $n>1$. This crucial finding implies that, as
the variable $x$ becomes critical for $g=1/a_2$, all the other
baby universe variables are still away from criticality.
Since their propagators are massive,
their fate is the same as of the variable $v$ in the Ising case:
integrating them out makes no effect on the relevant terms in the
effective action for $x$.
For $g=1/a_2$, the important integral
over $x$ reduces to
$$\bar F(\Delta, N^2)=\log
\int_{-\infty}^\infty dx N e^{2\pi R N^2 \Delta x + F_0(x, N^2) }
\eqn\eq$$
where $\Delta= a_2 (c_2 -{a_1\over a_2}-\lambda)$.
While in other models we could express this integral directly in terms of
scaling variables, for $c=1$ this is impossible because of the
logarithmic scaling violations in $F_0(x, N^2)$. Actually, the situation
turns out to be even simpler than for $c<1$.

In the leading saddle point approximation,
$$\eqalign{&\bar F(\Delta, N^2)= 2\pi R N^2 \Delta x_s +  F_0(x_s, N^2)\
,\cr
& -{\partial F_0\over\partial x} (x=x_s)= 2\pi R N^2 \Delta \ .}
\eqn\exactsp$$
Thus, $\bar F$ is the Legendre transform of $-F_0$, with
$2\pi R N^2 \Delta$ being the conjugate variable of $x$.
The leading order relation between $\Delta$ and $x$ is
$$ \Delta\log\Delta\sim x\ , $$
in agreement with ref. [\SI]. Our analysis of integration around the
saddle point, based on eq. \saddle, indicates that, remarkably, all such
corrections are suppressed by powers of $\log\Delta$. Thus,
in the double scaling limit where $N\Delta$ is kept fixed, eq.
\exactsp\ is exact.  This Legendre transform was introduced in
ref. [\GK] to calculate the sum over ``one puncture irreducible''
surfaces. It was shown to satisfy a simple equation,
$$\eqalign{&{\partial^2 \bar F\over\partial \Delta^2}=
2\pi R N^2 \tilde\rho(\Delta)=R N^2
\left [-\ln\Delta+\sum_{m=1}^\infty\left ( 2 N\Delta\sqrt R\right )^{-2m}
f_m(R)\right ],\cr
&f_m(R)=(2m-1)! \sum_{k=0}^m |2^{2k}-2|~|2^{2(m-k)}-2|
{|B_{2k}|~|B_{2(m-k)}|\over (2k)! [2(m-k)]!} R^{m-2k}\ ,
}\eqn\db $$
where $\tilde\rho(\Delta)$ is the temperature corrected density of
states, and $\Delta$ is the distance of the Fermi
level from the top of the potential.
Integrating eq. \db, we find
$$\bar F=
{1\over 8}\left\{-
(2N\Delta\sqrt R)^2\ln\Delta-2f_1(R)\ln\Delta+
\sum_{m=1}^\infty {f_{m+1}(R)\over m(2m+1)}
(2N\Delta\sqrt R)^{-2m}\right\}\ , \eqn\radexp$$
where we have exhibited only the terms that survive in the
double-scaling limit. It is remarkable that the $c=1$ model with fine-tuned
wormhole weights directly generates the ``one puncture irreducible''
surfaces.
This model, which has no scaling violations as a function of the area [\SI],
is in many ways simpler and more natural than the
conventional $c=1$ model.

\chapter{New Gravitational Dimensions}

One lesson we can draw from the preceding
section is that, even though a given model
may have many baby universe integration variables, it is usually the case
that only the integral over the lowest dimension coupling affects the
critical behavior. It is clear, however, that this cannot be
the most general situation.
In this section we show how to make other integrations relevant by
changing the type of trace-squared terms added
to the action.

As an instructive example, let us consider the modified Ising model of
section 2.2 with a more general class of trace-squared terms,
$$ {g\over 2}\left ( \Tr (\Phi_1^4+\Phi_2^4) \right)^2 + {g'\over 2}
\left( \Tr (\Phi_1^4-\Phi_2^4) \right)^2
\ .\eqn\tsq$$
For $g\neq g'$ this introduces a term of the form $\Tr\Phi_1^4
\Tr\Phi_2^4$ which generates wormholes with opposite values of spin at
the two ends. It is not surprising that such processes can make the
integration over the spin field coupling constant relevant. We may,
for instance, set $g=0$ (actually, any $g<1/a_2$ will do), while
fine tuning $g'$ to its critical value.
The partition function becomes
$$ \eqalign{& Z\sim
\int_{-\infty}^\infty dv e^{-{N^2 v^2\over 2g'}}
\int {\cal D} \Phi_1 {\cal D} \Phi_2 \, e^{-N S}\ , \cr
& S =
   \Tr \left[ \Phi_1^2+\Phi_2^2+ k\Tr \Phi_1 \Phi_2-
     \lambda ( \Phi_1^4+ \Phi_2^4)- (v+\tau_\sigma)
 ( \Phi_1^4- \Phi_2^4) \right]
}\eqn\newisingtrick$$
where we have introduced coupling constant $\tau_\sigma$
in order to study correlation functions of $\Tr (\Phi_1^4-\Phi_2^4)$.
Defining a shifted variable $u=v+\tau_\sigma$, we perform the matrix
integral first and reduce the modified free energy to
$$\eqalign{& \bar F=  F_0(t) + \log \int du e^{f(u)} \ ,\cr
& f(u)= -{N^2\over 2 g'} (u^2- 2u \tau_\sigma + \tau_\sigma^2)+
F_1 (\Delta, u, N^2)\ ,
\cr
& F_1 =\half N^2 u^2 (b+b'\Delta^{2d-\gamma}+\ldots) +
N^2 u^4 b'' \Delta^{4d-2-\gamma} + \ldots
}\eqn\eq
$$
where $t\sim \Delta N^{2/(2-\gamma)}$ and $\Delta=c_2-\lambda$.
The universal part of $F_1$ is a function of $t$ and
$t_\sigma= u N^{2(1-d)/(2-\gamma)}$. In our specific case
$\gamma=-1/3$ and $d=1/6$. If we now fine tune $g'=1/b$ and introduce
the scaling variable
$$\eqalign{&
\bar t_\sigma= b\tau_\sigma N^{2(1-\bar d)/(2-\gamma)}\ , \cr
&\bar d= \gamma-d\ , }\eqn\newdimension
$$
then the universal part of the modified free energy is given by
$$\bar F (t,\bar t_\sigma) =\log \int_{-\infty}^\infty dt_\sigma
e^{t_\sigma \bar t_\sigma + F(t, t_\sigma)}
\eqn\relsig$$
Here $F(t, t_\sigma)$ is the universal part of the conventional sum over
surfaces.

Eq. \newdimension\ implies that the gravitational dimension of the spin
field has changed from $d=1/6$ to $\bar d=-1/2$. Thus, in modified
\mm s negative dimensions arise naturally. Although we have discussed
a specific example, the change of gravitational dimension given by
eq. \newdimension\ is general. As shown in section 1.1, this formula
agrees with the change in dimension caused by changing the branch
of \L\ dressing. Therefore, there are serious reasons to believe that
such operators, which were previously thought not to exist, are in
fact present in the spectra of modified \mm s.

Eq. \relsig\ shows that, by a fine-tuning of $g'$, the coupling
constant
corresponding to the spin field has been driven to criticality.
It is not hard to see that a simultaneous tuning of $g$ to $1/a_2$
also makes the coupling constant $t$, corresponding to the puncture operator,
critical, so that
$$\bar F (\bar t,\bar t_\sigma) =\log \int_{-\infty}^\infty  dt dt_\sigma
e^{t \bar t +t_\sigma \bar t_\sigma + F(t, t_\sigma)}
\ .\eqn\eq$$
It is now clear that a fine-tuning of $n$ parameters in the
trace-squared terms can result in integration over $n$ coupling
constants, giving the general formula \gendouble.

\chapter{Sum over Surfaces of Genus One}

In this section we focus on the torus contribution to the free energy
which, in any string theory, is directly related to the spectrum.
For all conventional matrix models the torus free energy has been
successfully reproduced by path integration in \L\ theory [\BK,\st].
For this reason it is particularly interesting to calculate the
corresponding quantity in modified \mm s and ask for its continuum
interpretation.

For any $(p, q)$ minimal model coupled to gravity the
torus free energy is\foot{We are quoting the answer for \mm s
with non-symmetric potentials. For corresponding
models with symmetric potentials the free energy is doubled.}
$$ F^{G=1} (t) = -{(p-1)(q-1)\over 24 (p+q-1)}\log t
\ .\eqn\convtor$$
This result was reproduced [\BK] in \L\ theory with the interaction
term of eq. \usint. Let us now study the modifications to this result
due to the integration over coupling constants. Consider, for instance,
a model where the gravitational dimension of operator $O$ has
been changed from $d_O$ to $\gamma-d_O$.
Here the modified sum over surfaces is given by eq. \newdouble.
Setting $\bar t_O=0$ and performing gaussian integration around the
saddle point, we find
$$ \bar F^{G=1} (t)= F^{G=1} (t)+\half (\gamma-2 d_O) \log t
\ .\eqn\corrtor$$
This result is puzzling from the point of view of the simplest
\L\ approach. Since we have not changed the string susceptibility
exponent, it would seem that the \L\ action is still \usint, and that
the calculation of ref. [\BK] with the result \convtor\ should still
apply. The \mm\ tells us otherwise: the moment we change the
gravitational dimension of an operator, the torus free energy receives
a correction. We may speculate that in \L\ theory this correction
originates from a boundary term in the modular integral, but at the
moment we do not know how to derive it directly. In the following,
however, we will give a plausibility argument for the presence of
the correction found in eq. \corrtor.

Our argument is based on the interpretation [\BKN] of the torus
free energy in $(p, q)$ models as the sum over zero-point energies
of an infinite number of one-dimensional particles
(harmonic oscillators). Each oscillator
corresponds to an operator in \L\ theory of the form
$\CO e^{\beta\phi}$, where $\CO$ has dimension $h$ and
$$\eqalign{&\beta=-{Q\over 2}+\omega\ , \cr
&\omega= \sqrt {{Q^2\over 4} - 2 + 2h}\ .}
$$
$\omega$, which is the ``\L\ energy'', gives the frequency of the
oscillator. {\it A priori} there is a sign ambiguity for $\omega$,
but in conventional \L\ theory all $\omega$ are taken to be positive.
Each operator contributes zero-point energy $\half\omega$ to the
coefficient of the \L\ volume, $-\log t/\alpha$, where
$\alpha=\alpha_+$ from eq. \usint. Thus,
$$ F^{G=1} (t)= -{\log t\over\alpha}\sum_i {\half\omega_i}
\eqn\zprep$$
For the $(p, q)$ model the spectrum of energies is given by [\BKN]
$$ \omega_i={s_i\over\sqrt{2pq}}
$$
where $s_i$ are the positive integers not divisible by either $p$ or
$q$. After substituting this into \zprep\ and using zeta function
regularization of the infinite sum, ref. [\BKN] recovered eq. \convtor.

When we fine tune the theory as in section 4, we replace $\omega_O$ by
$-\omega_O$ in the \L\ dressing of operator $O$. If we make the same
replacement in eq. \zprep, we arrive at
$$ \bar F^{G=1} (t)= F^{G=1} (t)+{|\omega_O|\over\alpha} \log t
\ .\eqn\prediction
$$
Remarkably, since $\gamma-2 d_O=2{|\omega_O|\over\alpha}$,
this agrees with the matrix model result, eq. \corrtor! This suggests
a connection between the operator content and the torus free energy
in modified \mm s. Our basic premise is that a fine-tuned operator
with negative \L\ energy, $-|\omega|$, contributes a negative zero-point
energy, $-\half |\omega|$. This suggests that, from the space-time point
of view, a negatively dressed operator is a fermion.
It would be interesting to find an explanation for this effect.

Proceeding to other modified \mm s we note that, for the models
described by relation \double,
$$\bar F^{G=1} (\bar t) = -{(p-1)(q-1)\over 24 (p+q+1)}\log \bar t
-{1\over p+q+1} \log \bar t
\ .\eqn\ctor$$
A naive \L\ calculation with potential \modpot\ would give only
the first term in the above [\I]. However, applying eq. \zprep\ with
$\alpha=\alpha_-$ from eq. \modpot, and including
negative zero-point energy for $O_{\rm min}$,
reproduces the \mm\ result.

The second terms in eqs. \corrtor\ and \ctor\ are due to integration
around the saddle point in eqs. \newdouble\ and \double. These
corrections can be eliminated if we treat the baby universe variables
classically, \ie\ if we freeze them at their saddle point values. In
such a theory
$\bar F$ would simply be the Legendre transform of $-F$.
Of course, this theory does not correspond to the original \mm\ with
trace-squared terms, but it does have a simple geometric interpretation.
It calculates the sum over trees of touching
random surfaces (bubbles), with each bubble allowed to have
arbitrary genus. In other words, the wormholes are present, but they
are not allowed to increase the overall genus. Since from the world
sheet point of view this constraint is highly non-local, we do not
regard such a theory as natural. Eq. \ctor\ shows, however, that the
relative importance of surfaces where a wormhole closes the loop
decreases with increasing $p$ and $q$. For $c=1$ the quantum effects
associated with the baby universe variables become completely negligible,
so that the modified sum over surfaces is simply the Legendre transform,
eq. \exactsp.

\chapter{Conclusions}

Our non-perturbative solutions of \mm s modified by various trace-squared
terms strongly suggest that there exists a continuum \L\ formulation of
these models. All the modified scaling exponents agree with the idea [\I]
that a fine-tuning of trace-squared terms changes the branch of
\L\ dressing of some operators. Since the new branch of dressing does not
have a semiclassical limit, the resulting \L\ theory is more complicated and
more interesting than the conventional one.
We hope that our \mm\ results
will provide a useful guide towards such a theory.

The solution of the modified \mm s is also
quite interesting in itself because of
its connection with general wormhole phenomena in \qg.
The microscopic wormholes, introduced by the trace-squared terms,
lead to integration over coupling constants, as expected on general grounds.
Such integration arises in any theory with bilocal operators in the
action. Physical effects of integration over coupling constants
have even been found in theory of elasticity, where
they change the order of phase transitions\foot{We are grateful
to A. Polyakov for telling us about this effect.} [\lp].
Our work provides another example of a system where integration
over coupling constants introduces a profound change, affecting even the order
of the phase transition for planar surfaces.
It is interesting to look for other physical systems where coupling
constants turn into dynamical variables.

\ack
We are indebted to V. Periwal for very helpful suggestions.
We are also grateful to K. Demeterfi, D. Gross and A. Polyakov for
useful discussions.
This work was supported in part by DOE grant DE-FG02-91ER40671,
the NSF Presidential Young Investigator Award PHY-9157482,
James S. McDonnell Foundation grant No. 91-48,
and an A. P. Sloan Foundation Research Fellowship.

\refout
\bye